\RequirePackage[2020-02-02]{latexrelease}

\documentclass[twocolumn,preprintnumbers]{revtex4}%
\usepackage{amssymb}
\usepackage{amsmath}
\usepackage{graphicx}
\usepackage{epstopdf}
\usepackage{dcolumn}
\usepackage{bm}
\usepackage{xcolor}
\usepackage{ifthen}
\usepackage[normalem]{ulem}
\usepackage{amsfonts}%
\RequirePackage[2020-02-02]{latexrelease}
\graphicspath{{c:/Users/Eyal/MyDocuments/latex/Texfiles/Diamond/dipolarBA/EntangSupp/DISSB/}
	{C:/Users/Eyal/MyDocuments/latex/Texfiles/Diamond/dipolarBA/EntangSupp/Q/}
	{C:/Users/Eyal/MyDocuments/latex/Texfiles/Diamond/dipolarBA/EntangSupp/SD/}
	{C:/Users/Eyal/MyDocuments/latex/Texfiles/Diamond/dipolarBA/EntangSupp/PP/}}
\UseRawInputEncoding
\providecommand{\U}[1]{\protect\rule{.1in}{.1in}}
\providecommand{\U}[1]{\protect\rule{.1in}{.1in}}
\def\showal{1}
\newcommand{\al}[1]{\ifthenelse{\showal=1}{\textcolor{orange}{[[#1]]}}{}}

\newcommand{\eb}[1]{\ifthenelse{\showal=1}{\textcolor{cyan}{[[#1]]}}{}}
\begin{document}
\title{The spontaneous disentanglement hypothesis and causality}
\author{Eyal Buks}
\email{eyal@ee.technion.ac.il}
\affiliation{Andrew and Erna Viterbi Department of Electrical Engineering, Technion, Haifa
32000, Israel}
\date{\today }

\begin{abstract}
The hypothesis that disentanglement spontaneously occurs in quantum systems is
motivated by some outstanding issues in the foundations of quantum mechanics.
However, for some cases, spontaneous disentanglement enables the violation of
the causality principle. To mitigate the conflict with causality, a
formulation for the hypothesis, which is based on the maximum entropy
principle, is proposed. The method of Lagrange multipliers is implemented to
ensure consistency with causality. The proposed formulation is applicable for
any quantum system having a Hilbert space of finite dimensionality.

\end{abstract}
\maketitle





\textbf{Introduction} -- Unitary time evolution of a quantum state vector is
governed by the Schr\"{o}dinger equation. Standard quantum mechanics (QM) is
commonly formulated based on the assumption that the unitary time evolution is
supplemented by two additional auxiliary processes, which are both not
derivable from the Schr\"{o}dinger equation. The first one is a collapse of
the state vector, which occurs when a measurement is performed, and the second
one is thermalization. As was first shown by Schr\"{o}dinger
\cite{Schrodinger_807}, the process of collapse gives rise to an internal
inconsistency \cite{Penrose_4864}, which became known as the problem of
quantum measurement. The conflict between the time--reversibility of the
Schr\"{o}dinger equation and the time--irreversibility of the process of
thermalization is commonly referred to as the arrow of time problem.

Unitary time evolution in standard QM is governed by linear equations of
motion. Both the measurement and the arrow of time problems have motivated the
study of a variety of nonlinear extensions to QM
\cite{Weinberg_61,Grabert_161,Doebner_3764,Gisin_5677,Gisin_2259,Kaplan_055002,Munoz_110503,Geller_2200156,Ghirardi_470,Oppenheim_041040,Bennett_170502,Czachor_4122}%
. Some of the proposed nonlinear extensions yield a spontaneous collapse of
the state vector
\cite{Bassi_471,Pearle_857,Kowalski_1,Fernengel_385701,Carlesso_243,Donadi_74}%
. The collapse gives rise to disentanglement between a quantum system and its
measuring apparatus. The generation of disentanglement, however, requires
nonlinearity, because the subset of fully disentangled states within the
Hilbert space of a given multipartite quantum system is generally not a linear
subspace, and consequently, disentanglement cannot be generated by equations
of motion that obey the superposition principle \footnote{To demonstrate why nonlinearity is needed to generate
disentanglement, consider, as an example, a two spin system. The states
$\left\vert \uparrow\uparrow\right\rangle $ and $\left\vert \downarrow
\downarrow\right\rangle $ are both fully disentangled. The requirement that
disentanglement has no impact on product (i.e. disentangled) states, together
with the assumption that the equations of motion satisfy the superposition
principle, lead to the conclusion that all states having the form
$\alpha\left\vert \uparrow\uparrow\right\rangle +\beta\left\vert
\downarrow\downarrow\right\rangle $, where $\left\vert \alpha\right\vert
^{2}+\left\vert \beta\right\vert ^{2}=1$, are unaffected by disentanglement.}. Moreover, thermalization, which can be described
as an entropy maximization process \cite{Presse_1115}, requires nonlinearity,
because the entropy $\sigma=\left\langle -\log\rho\right\rangle $ is a
nonlinear function of the density operator $\rho$
\cite{Grabert_161,Ottinger_026121}. Note that multi--stabilities in finite
systems, which are excluded by standard QM \cite{Callender_539}, become
possible provided that nonlinearity is permitted \cite{Buks_012439}.

For some cases, however, nonlinear quantum dynamics may give rise to conflicts
with well-established physical principles, such as separability
\cite{Hejlesen_thesis,Jordan_022101,Jordan_012010} and causality
\cite{Bassi_055027,Polchinski_397,Helou_012021,Rembielinski_012027,Rembielinski_420,Rembielinski_2605_13756,Aharonov_3316,Polo_2506_18906,Oskeka_2512_19260,Ray_1018}%
. The conflict with Einstein's causality principle is commonly demonstrated by
showing that nonlinear dynamics enable superluminal (i.e. faster than light) signaling.

Note that, even in standard QM, where unitary time evolution is
linear, a conflict with causality cannot be fully avoided (e.g. see
\cite{Aharonov_359,Chiao_345}). The vacuum speed of light $c$ does not appear
in the Schr\"{o}dinger equation, and consequently, a speed limit related to
$c$ cannot be derived from standard, and non--relativistic, QM. The
relativistic version of QM partially addresses some of these difficulties.
However, it has been shown that the Dirac equation, similarly to the
Schr\"{o}dinger one, can give rise to superluminal tunneling (e.g. see
\cite{Krekora_032107}).

In his seminal paper \cite{Gisin_1}, Gisin has
explored the conflict with causality for a quantum entangled state. He found
that, under some conditions, superluminal signaling becomes possible, provided
that quantum time evolution is allowed to be nonlinear. An example protocol to realize the Gisin's superluminal telegraph is described
below. Consider a system composed of two subsystems, labeled as A and B,
respectively \cite{Diosi_012019}. It is assumed that nonlinearity stems from
the process of spontaneous disentanglement \cite{Buks_e00986}. Alice owns
subsystem A, which is a \textit{single} qubit, whereas Bob owns subsystem B,
which is a \textit{pair} containing two qubits. The state of the entire system
$\left\vert \psi\right\rangle $ is assumed to be a Greenberger Horne Zeilinger
(GHZ) state given by (subscripts A and B refer to Alice and Bob, respectively)%
\begin{align}
\left\vert \psi\right\rangle  &  =\frac{\left\vert \uparrow\right\rangle
_{\mathrm{A}}\otimes\left\vert \uparrow\uparrow\right\rangle _{\mathrm{B}%
}-\left\vert \downarrow\right\rangle _{\mathrm{A}}\otimes\left\vert
\downarrow\downarrow\right\rangle _{\mathrm{B}}}{\sqrt{2}}\nonumber\\
&  =\frac{\left\vert \longrightarrow\right\rangle _{\mathrm{A}}\otimes
\left\vert -\right\rangle _{\mathrm{B}}+\left\vert \longleftarrow\right\rangle
_{\mathrm{A}}\otimes\left\vert +\right\rangle _{\mathrm{B}}}{\sqrt{2}%
}\;.\nonumber\\
&
\end{align}
The arrow symbols $\uparrow$, $\downarrow$, $\longrightarrow$ and
$\longleftarrow$ label eigenvectors of the matrix $\boldsymbol{\sigma}\cdot
\mathbf{\hat{u}}$, with an eigenvalue of $+1$, where $\boldsymbol{\sigma
}=\left(  \sigma_{1},\sigma_{2},\sigma_{3}\right)  $ is the Pauli spin matrix
vector, and the unit vector $\mathbf{\hat{u}}$ is pointing in the
$\mathbf{\hat{z}}$, $-\mathbf{\hat{z}}$, $\mathbf{\hat{x}}$ and $-\mathbf{\hat
{x}}$, directions, respectively. Since both $\left\vert \uparrow
\uparrow\right\rangle _{\mathrm{B}}$ and $\left\vert \downarrow\downarrow
\right\rangle _{\mathrm{B}}$ are product states, disentanglement
\textit{within} subsystem B has no effect (disentanglement between subsystems
A and B is disregarded, for simplicity). However, if Alice performs a quantum
measurement in the basis $\left\{  \left\vert \longrightarrow\right\rangle
_{\mathrm{A}},\left\vert \longleftarrow\right\rangle _{\mathrm{A}}\right\}  $,
where $\left\vert \longrightarrow\right\rangle _{\mathrm{A}}=\left(
\left\vert \uparrow\right\rangle _{\mathrm{A}}+\left\vert \downarrow
\right\rangle _{\mathrm{A}}\right)  /\sqrt{2}$ and $\left\vert \longleftarrow
\right\rangle _{\mathrm{A}}=\left(  \left\vert \uparrow\right\rangle
_{\mathrm{A}}-\left\vert \downarrow\right\rangle _{\mathrm{A}}\right)
/\sqrt{2}$, the state of subsystem B is expected to collapse into the state
$\left\vert +\right\rangle _{\mathrm{B}}$, with probability 1/2, or into the
state $\left\vert -\right\rangle _{\mathrm{B}}$,with the same probability of
1/2, where $\left\vert \pm\right\rangle _{\mathrm{B}}=\left(  \left\vert
\uparrow\uparrow\right\rangle _{\mathrm{B}}\pm\left\vert \downarrow
\downarrow\right\rangle _{\mathrm{B}}\right)  /\sqrt{2}$. Both states
$\left\vert +\right\rangle _{\mathrm{B}}$ and $\left\vert -\right\rangle
_{\mathrm{B}}$ are fully entangled, and thus a disentanglement process
occurring after Alice's measurement, is expected to give rise to an impact
that can be detected by Bob (for example, by measuring the expectation values
of Bell operators for the pair he owns). This detectability arguably enables
superluminal signaling.

For the same above--discussed protocol, predictions that are derived from the
spontaneous disentanglement hypothesis are entirely different. For example,
according to this hypothesis, rather than performing measurements, all Alice
and Bob can do is turn on and off coupling between the quantum system they own
and measuring apparatuses. Moreover, no collapse occurs within the framework
of this hypothesis. Nevertheless, for some cases, disentanglement of a
spatially extended quantum system may give rise to a conflict with the
causality principle. In the current study, this conflict is explored, and a
way to avoid it is proposed.

\textbf{The collapse postulate} -- Unitary time evolution is
linear according to standard QM. This linearity is one of the assumptions that
are commonly used to show that superluminal signaling is excluded. However, as
is discussed below, another important assumption is related to the way the
collapse postulate is formulated. Consider a bipartite system, composed of a
quantum subsystem, which is labeled by the letter a, and a measuring
apparatus, which is labeled by the letter b. The subsystems' density operators
$\rho_{\mathrm{a}}$ and $\rho_{\mathrm{b}}$\ are derived from the density
operator $\rho$ of the composed bipartite system by partial tracing, i.e.
$\rho_{\mathrm{a}}=\operatorname{Tr}_{\mathrm{b}}\rho$ and $\rho_{\mathrm{b}%
}=\operatorname{Tr}_{\mathrm{a}}\rho$.

In standard QM, a measurement is described as a process having two steps. In
the first one, a pure entangled state $\left\vert \psi\right\rangle $ is
generated by the unitary time evolution that is induced by the coupling
between the quantum system and its measuring apparatus. The Schmidt
decomposition can be applied to express the pure state $\left\vert
\psi\right\rangle $ as%
\begin{equation}
\left\vert \psi\right\rangle =\sum_{s}p_{s}^{1/2}\left\vert s\right\rangle
_{\mathrm{a}}\otimes\left\vert s\right\rangle _{\mathrm{b}}\;,
\label{Schmidt decomposition}%
\end{equation}
where $p_{s}\in\left[  0,1\right]  $, $\sum_{s}p_{s}=1$, and where both vector
sets $\left\{  \left\vert s\right\rangle _{\mathrm{a}}\right\}  $ and
$\left\{  \left\vert s\right\rangle _{\mathrm{b}}\right\}  $ are orthonormal.
In the Schmidt basis, the reduced density operators are given by
$\rho_{\mathrm{a}}=\sum_{s}p_{s}\left\vert s\right\rangle _{\mathrm{a}%
\;\mathrm{a}}\left\langle s\right\vert $ and $\rho_{\mathrm{b}}=\sum_{s}%
p_{s}\left\vert s\right\rangle _{\mathrm{b}\;\mathrm{b}}\left\langle
s\right\vert $. According to the collapse postulate, under some unspecified
conditions, the density operator, which prior to the collapse is given by
$\rho=\left\vert \psi\right\rangle \left\langle \psi\right\vert $, undergoes
an abrupt change.

The measurement problem stems from the difficulty to self-consistently specify
triggering conditions for the collapse process. Another important question is
what is the transformation that is applied to the density operator $\rho$ due
to a collapse. When one of the subsystems, or both, are spatially extended,
the collapse generally may give rise to a conflict with the causality
principle. However, such a conflict can be avoided, provided that the collapse
conserves both $\rho_{\mathrm{a}}$ and $\rho_{\mathrm{b}}$. Clearly, this
requirement, that both $\rho_{\mathrm{a}}$ and $\rho_{\mathrm{b}}$ are kept
unchanged, does not uniquely determine the transformation associated with the
collapse process. However, this requirement is sufficient to yield the Born
rule \cite{Vaidman_567,Mukherjee_20250254,Mukherjee_2410_08844,Gisin_363}.
Moreover, it ensures that superluminal signaling is excluded, since the
collapse has no impact on any subsystem's property
\cite{Kent_012108,Emilia_053002,Czachor_139,Aleksander_245014}.

The spontaneous disentanglement hypothesis is formulated using nonlinear
equations of motion, which are described in the next section. In the following
section, a method to mitigate the conflict with the causality principle, which
is based on the same assumption, that the reduced density operators
$\rho_{\mathrm{a}}$ and $\rho_{\mathrm{b}}$ are kept fixed, is proposed.

\textbf{Nonlinear extension} -- Consider a modified master
equation for the density operator $\rho$ given by%
\begin{equation}
\frac{\mathrm{d}\rho}{\mathrm{d}t}=i\hbar^{-1}\left[  \rho,\mathcal{H}\right]
+\Omega\left(  \Theta\right)  \;, \label{modified master equation}%
\end{equation}
where $\hbar$ is the reduced Planck's constant, $\mathcal{H}^{{}}%
=\mathcal{H}^{\dag}$ is the system's Hamiltonian, the positive semi--definite
operator $\Theta$ is allowed to depend on $\rho$, for a general operator $X$,
the operator $\Omega\left(  X\right)  $\ is defined by $\Omega\left(
X\right)  =-X\rho-\rho X+2\left\langle X\right\rangle \rho$, and $\left\langle
X\right\rangle =\operatorname{Tr}\left(  X\rho\right)  $. As is shown in
appendix \ref{AppK}, the master equation (\ref{modified master equation}) can
be derived from nonlinear Kraus operators \cite{Elben_200501,Sergi_1350163,Brody_230405,Kaplan_055002,Geller_2200156},
and from a stochastic Langevin--Schr\"{o}dinger equation for the state vector
given by Eq. (\ref{SLE}) of appendix \ref{AppK}. The modified master equation
(\ref{modified master equation}) conserves both norm [note that $\left(
\mathrm{d}/\mathrm{d}t\right)  \operatorname{Tr}\rho=0$, provided that
$\operatorname{Tr}\rho=1$] and positivity \cite{Gorini_821} [see Eq. (2.202)
of Ref. \cite{Buks_QMLN}] of $\rho$.

The Langevin--Schr\"{o}dinger equation (\ref{SLE}) (see appendix \ref{AppK}), which contains a term
proportional to $\left\langle \Theta\right\rangle ^{1/2}$, can be derived from
a modified master equation having a given $\Omega\left(  \Theta\right)  $ term
[see Eq.(\ref{modified master equation})], provided that the mapping
$\Omega\left(  X\right)  =-X\rho-\rho X+2\left\langle X\right\rangle \rho
$\ can be inverted. Any bounded operator $X$ can be decomposed as
$X=X_{0}+X_{1}$, where $X_{1}$ is traceless, and $X_{0}$, which is given by
$X_{0}=\left(  d^{-1}\operatorname{Tr}X\right)  I$, is proportional to the
identity operator $I$, where $d$ is the dimensionality (note that
$\operatorname{Tr}X_{0}=\operatorname{Tr}X_{{}}$). The following holds
$\Omega\left(  X_{{}}\right)  =\Omega\left(  X_{1}\right)  $, and thus the
mapping $\Omega\left(  X\right)  $\ is generally not invertible [i.e. a given
operator $\Omega\left(  X\right)  $\ does not uniquely determine the operator
$X$, since $\Omega\left(  X_{{}}\right)  $ is independent of $X_{0}$]. To
allow invertibility, the mapping $\Omega\left(  X\right)  $ is replaced by
$\tilde{\Omega}\left(  X\right)  $, which is given by $\tilde{\Omega}\left(
X\right)  =\Omega\left(  X\right)  +I\operatorname{Tr}X$, and for which the
following holds $\tilde{\Omega}\left(  X\right)  =\Omega\left(  X_{1}\right)
+I\operatorname{Tr}X_{0}$. The mapping $\tilde{\Omega}\left(  X\right)  $ is
invertible if and only if $\dim\left(  \ker\rho\right)  =0$, as can be shown
by expressing $\rho$ in terms of its normalized eigenvectors $\left\vert
\phi_{n}\right\rangle $ and eigenvalues $r_{n}\in\left[  0,1\right]  $ as
$\rho=\sum_{n}r_{n}\left\vert \phi_{n}\right\rangle \left\langle \phi
_{n}\right\vert $, and by expressing $X$ in the same basis as $X=\sum
_{n,m}x_{n,m}\left\vert \phi_{n}\right\rangle \left\langle \phi_{m}\right\vert
$, where $x_{n,m}=\left\langle \phi_{n}\right\vert X\left\vert \phi
_{m}\right\rangle $. For the case where $\dim\left(  \ker\rho\right)  >0$,
i.e. $r_{n}=0$ for some $n$, the following holds $\Omega\left(  \left\vert
\phi_{n}\right\rangle \left\langle \phi_{n}\right\vert \right)  =0$, thus
$\tilde{\Omega}\left(  \left\vert \phi_{n}\right\rangle \left\langle \phi
_{n}\right\vert -d^{-1}I\right)  =0$, hence $\tilde{\Omega}\left(  X\right)  $
is not invertible. On the other hand, for the case where $\dim\left(  \ker
\rho\right)  =0$, i.e. $r_{n}>0$ for all $n$, the mapping $\tilde{\Omega
}\left(  X\right)  $ is invertible, since for this case the only solution for
$\tilde{\Omega}\left(  X\right)  =0$ is $X=0$ [note that $\tilde{\Omega
}\left(  X\right)  =0$ implies that $\operatorname{Tr}X=0$ and $\left(
r_{m}+r_{n}\right)  x_{n,m}=2\left\langle X\right\rangle r_{n}\delta_{n,m}$].

For a general explicitly time--independent observable $A$, the modified master
equation (\ref{modified master equation}) yields%
\begin{equation}
\frac{\mathrm{d}\left\langle A\right\rangle }{\mathrm{d}t}=-i\hbar
^{-1}\left\langle \left[  A,\mathcal{H}\right]  \right\rangle -\left\langle
\Delta_{A}\Delta_{\Theta}+\Delta_{\Theta}\Delta_{A}\right\rangle \;,
\label{d<A>/dt}%
\end{equation}
where $\Delta_{A}=A-\left\langle A\right\rangle $ and $\Delta_{\Theta}%
=\Theta-\left\langle \Theta\right\rangle $. Note that for the case where
$\mathcal{H}=0$ and $A=\Theta$, Eq. (\ref{d<A>/dt}) yields $\mathrm{d}%
\left\langle \Theta\right\rangle /\mathrm{d}t=-2\left\langle \Delta_{\Theta
}^{2}\right\rangle \leq0$ for a fixed $\Theta$. This result suggests that the
nonlinear term in the modified master equation (\ref{modified master equation}%
) gives rise to the suppression of the expectation value $\left\langle
\Theta\right\rangle $. Hence, the nonlinear term can be employed to suppress a
given physical property, provided that $\left\langle \Theta\right\rangle $
quantifies that property. As is discussed below, this suppression can be used
to generate both thermalization and disentanglement.

\textbf{Thermalization} -- According to Jaynes' principle
\cite{Jaynes_579,Presse_1115}, the density operator of a given system in
thermal equilibrium $\rho_{0}$ maximizes the entropy $\sigma=\left\langle
-\log\rho\right\rangle $ under some constraints. Alternatively, $\rho_{0}$ can
be derived by minimizing a free energy, which can be defined for any given set
of linear constraints. For example, when the energy expectation value
$\left\langle \mathcal{H}\right\rangle $ is constrained ($\mathcal{H}$, which
is assumed to be time--independent, is the Hamiltonian), the state of thermal
equilibrium $\rho_{0}$, which is given by $\rho_{0}=e^{-\beta\mathcal{H}%
}/\operatorname{Tr}\left(  e^{-\beta\mathcal{H}}\right)  $ [see Eq. (8.108) of
Ref. \cite{Buks_QMLN}], can be found by minimizing the Helmholtz free energy
$\left\langle \mathcal{U}_{\mathrm{H}}\right\rangle $, which is given by
$\left\langle \mathcal{U}_{\mathrm{H}}\right\rangle =\left\langle
\mathcal{H}\right\rangle -\beta^{-1}\sigma$. The Lagrange multiplier $\beta$
is given by $\beta=1/\left(  k_{\mathrm{B}}T\right)  $, where $k_{\mathrm{B}}$
is the Boltzmann's constant, and $T$ is the temperature. Thus, for this case,
thermalization can be generated by the nonlinear term in the modified master
equation (\ref{modified master equation}), provided that the operator $\Theta$
is taken to be proportional to the Helmholtz free energy operator
$\mathcal{U}_{\mathrm{H}}=\mathcal{H}+\beta^{-1}\log\rho$ \cite{Buks_e00986}.
As is discussed below, disentanglement can be obtained by combining Jaynes' maximum entropy principle with Einstein's causality.

\textbf{Subadditivity} -- The entropies of the quantum system $\sigma
_{\mathrm{a}}$, the measurement apparatus $\sigma_{\mathrm{b}}$, and the
bipartite composed system $\sigma_{{}}$ are given by $\sigma_{\mathrm{a}%
}=-\operatorname{Tr}\left(  \rho_{\mathrm{a}}\log\rho_{\mathrm{a}}\right)  $,
$\sigma_{\mathrm{b}}=-\operatorname{Tr}\left(  \rho_{\mathrm{b}}\log
\rho_{\mathrm{b}}\right)  $ and $\sigma=-\operatorname{Tr}\left(  \rho\log
\rho\right)  $, respectively. The relative entropy $\sigma\left(
\rho\parallel\rho_{\mathrm{a}}\otimes\rho_{\mathrm{b}}\right)  =\sigma
_{\mathrm{a}}+\sigma_{\mathrm{b}}-\sigma_{{}}$ quantifies the quantum mutual
information. It has been shown that relative entropy can be used to formulate
entanglement area laws \cite{Wolf_070502}, and both the second
\cite{Sagawa_125} and third \cite{Floerchinger_052117} laws of thermodynamics.

As was discussed above, the conflict with the causality principle can be
avoided, provided that the reduced density operators $\rho_{\mathrm{a}}$ (the
quantum system) and $\rho_{\mathrm{b}}$ (the measurement apparatus) are
unaffected by the process of disentanglement. According to the Klein
subadditivity inequality \cite{Klein_767,Wehrl_221}%
\begin{equation}
\sigma\leq\sigma_{\mathrm{a}}+\sigma_{\mathrm{b}}\;, \label{subadditivity}%
\end{equation}
and equality holds, i.e. $\sigma=\sigma_{\mathrm{a}}+\sigma_{\mathrm{b}}$, if
and only if $\rho=\rho_{\mathrm{a}}\otimes\rho_{\mathrm{b}}$. Hence
$\rho_{\mathrm{a}}\otimes\rho_{\mathrm{b}}$ maximizes the entropy $\sigma$,
under the constraints that both $\rho_{\mathrm{a}}$ and $\rho_{\mathrm{b}}$
are fixed. This observation suggests that the process of collapse can be
mimicked by the modified master equation (\ref{modified master equation}),
provided that the operator $\Theta$ is constructed such that the expectation
value $\left\langle -\Theta\right\rangle $ quantifies the total entropy
$\sigma$, and that the constraints that fix both $\rho_{\mathrm{a}}$ and
$\rho_{\mathrm{b}}$ are enforced. According to this approach, when
the constrains are disregarded, time evolution is governed by the Jaynes'
maximum entropy principle \cite{Jaynes_579}, which is replaced by a mutual
information minimum principle, when the constrains are enforced. These constraints can be described in terms of the generalized
Bloch vectors, which are defined in the next section.

\textbf{The Bloch matrix and vectors} -- Consider a bipartite
system composed of subsystems labeled by the letters a and b.
Let $D_{\mathrm{a}}$, $D_{\mathrm{b}}$ and $D_{\mathrm{H}}=D_{\mathrm{a}%
}D_{\mathrm{b}}$ be the Hilbert space dimensionality of subsystem a, subsystem
b, and the composed system, respectively (it is assumed that $D_{\mathrm{a}}$,
$D_{\mathrm{b}}$ and $D_{\mathrm{H}}$ are all finite). The generalized
Gell-Mann set $\left\{  \lambda_{l}\right\}  $, which spans the
SU($D_{\mathrm{H}}$) Lie algebra, contains $D_{\mathrm{H}}^{2}-1$ square
$D_{\mathrm{H}}\times D_{\mathrm{H}}$ Hermitian matrices. For the case
$D_{\mathrm{H}}=2$ ($D_{\mathrm{H}}=3$), the $D_{\mathrm{H}}^{2}-1=3$
($D_{\mathrm{H}}^{2}-1=8$) elements are called Pauli (Gell-Mann) matrices. The
Generalized Gell-Mann matrices are traceless, i.e. $\operatorname{Tr}%
\lambda_{l}=0$, and they satisfy the orthogonality relation $\left(
1/2\right)  \operatorname{Tr}\left(  \lambda_{l^{\prime}}\lambda
_{l^{\prime\prime}}\right)  =\delta_{l^{\prime},l^{\prime\prime}}$.

The generalized Gell-Mann $D_{\mathrm{L}}\times D_{\mathrm{L}}$ matrices
corresponding to subsystem $\mathrm{L}$, where $\mathrm{L}\in\left\{
\mathrm{a},\mathrm{b}\right\}  $, are denoted by $\lambda_{l}^{\left(
\mathrm{L}\right)  }$, where $l\in\left\{  1,2,\cdots,D_{\mathrm{L}}%
^{2}-1\right\}  $. Consider the set of $D_{\mathrm{H}}^{2}-1$ matrices
$G^{\left(  \mathrm{ab}\right)  }=\left\{  \Gamma_{a}^{\left(  \mathrm{a}%
\right)  }\otimes\Gamma_{b}^{\left(  \mathrm{b}\right)  }\right\}  -\left\{
\Gamma_{0}^{\left(  \mathrm{a}\right)  }\otimes\Gamma_{0}^{\left(
\mathrm{b}\right)  }\right\}  $, where $a\in\left\{  0,1,2,\cdots
,D_{\mathrm{a}}^{2}-1\right\}  $ and $b\in\left\{  0,1,2,\cdots,D_{\mathrm{b}%
}^{2}-1\right\}  $. For subsystem $\mathrm{L}$, where $\mathrm{L}\in\left\{
\mathrm{a},\mathrm{b}\right\}  $, the matrix $\Gamma_{0}^{\left(
\mathrm{L}\right)  }$ is defined by $\Gamma_{0}^{\left(  \mathrm{L}\right)
}=\left(  2^{1/4}/D_{\mathrm{L}}^{1/2}\right)  I_{\mathrm{L}}$,\ where
$I_{\mathrm{L}}$ is the $D_{\mathrm{L}}\times D_{\mathrm{L}}$ identity matrix,
and for $l\in\left\{  1,2,\cdots,D_{\mathrm{L}}^{2}-1\right\}  $\ the matrix
$\Gamma_{l}^{\left(  \mathrm{L}\right)  }$ is defined by $\Gamma_{l}^{\left(
\mathrm{L}\right)  }=2^{-1/4}\lambda_{l}^{\left(  \mathrm{L}\right)  }$.

With the help of the Kronecker matrix product identities $\operatorname{Tr}%
\left(  X_{1}\otimes X_{2}\right)  =\operatorname{Tr}X_{1}\operatorname{Tr}%
X_{2}$ and $\left(  X_{1}\otimes X_{2}\right)  \left(  X_{3}\otimes
X_{4}\right)  =\left(  X_{1}X_{3}\right)  \otimes\left(  X_{2}X_{4}\right)  $,
one finds that the set $G^{\left(  \mathrm{ab}\right)  }$ shares two
properties with the Gell-Mann set $\left\{  \lambda_{l}\right\}  $ of the
$D_{\mathrm{H}}$-dimensional Hilbert space. The first one is tracelessness
$\operatorname{Tr}G_{a,b}=0$ for any $G_{a,b}\equiv\Gamma_{a}^{\left(
\mathrm{a}\right)  }\otimes\Gamma_{b}^{\left(  \mathrm{b}\right)  }\in
G^{\left(  \mathrm{ab}\right)  }$ [recall that $G_{0,0}\notin G^{\left(
\mathrm{ab}\right)  }$], and the second one is orthogonality%
\begin{equation}
\frac{\operatorname{Tr}\left(  G_{a^{\prime},b^{\prime}}G_{a^{\prime\prime
},b^{\prime\prime}}\right)  }{2}=\delta_{a^{\prime},a^{\prime\prime}}%
\delta_{b^{\prime},b^{\prime\prime}}\;. \label{G OR}%
\end{equation}
The $D_{\mathrm{a}}^{2}\times D_{\mathrm{b}}^{2}$ matrix $B$, where
$B_{a,b}=\left\langle G_{a,b}\right\rangle $, is henceforth referred to as the
Bloch matrix \cite{Bertlmann_235303,Li_92}. The following holds $B_{0,0}%
=\sqrt{2/\left(  D_{\mathrm{a}}D_{\mathrm{b}}\right)  }$ (matrix elements'
numbering starts from $0$), and $\operatorname{Tr}\left(  B^{{}}B^{\dag
}\right)  =\operatorname{Tr}\left(  B^{\dag}B^{{}}\right)  =2\operatorname{Tr}%
\rho^{2}$ [see Eq. (\ref{G OR})]. The single subsystem Bloch vectors
$\mathbf{P}_{\mathrm{a}}=\left(  B_{1,0},B_{2,0},\cdots,B_{D_{\mathrm{a}}%
^{2}-1,0}\right)  $ and $\mathbf{P}_{\mathrm{b}}=\left(  B_{0,1}%
,B_{0,2},\cdots,B_{0,D_{\mathrm{b}}^{2}-1}\right)  $ are extracted from the
first column and first row, respectively, of the Bloch matrix $B$. The modified master equation (\ref{modified master equation}) yields equations
of motion for the entries of the Bloch matrix $B$ given by [see Eq.
(\ref{G OR})]%
\begin{align}
\frac{\mathrm{d}B_{a,b}}{\mathrm{d}t}  &  =i\hbar^{-1}\left\langle \left[
\mathcal{H},G_{a,b}\right]  \right\rangle +2\left\langle \Theta\right\rangle
B_{a,b}-\left\langle G_{a,b}\Theta+\Theta G_{a,b}\right\rangle \;.\nonumber\\
&
\end{align}
Note that the more general case of a multipartite system can be treated using
a Bloch tensor [see Eq. (8.411) of Ref. \cite{Buks_QMLN}].

The \textit{linear} relative entropy, which is denoted by
$\sigma_{\mathrm{LR}}$, is derived from the relative entropy $\sigma\left(
\rho\parallel\rho_{\mathrm{a}}\otimes\rho_{\mathrm{b}}\right)  $\ using the
approximation $\log x\simeq x-1$ \cite{Rahman_368}. It can be expressed in
terms of the entries of the Bloch matrix $B$ using the relation [see Eq.
(8.452) of Ref. \cite{Buks_QMLN}]%
\begin{equation}
\sigma_{\mathrm{LR}}=\sum_{a=1}^{D_{\mathrm{a}}^{2}-1}\sum_{b=1}%
^{D_{\mathrm{b}}^{2}-1}\frac{\left(  \left\langle \lambda_{a}^{\left(
\mathrm{a}\right)  }\otimes\lambda_{b}^{\left(  \mathrm{b}\right)
}\right\rangle -\left\langle \lambda_{a}^{\left(  \mathrm{a}\right)
}\right\rangle \left\langle \lambda_{b}^{\left(  \mathrm{b}\right)
}\right\rangle \right)  ^{2}}{4\left(  1-D_{\mathrm{\min}}^{-2}\right)  }\;,
\label{linear relative entropy S(rho)}%
\end{equation}
where $D_{\mathrm{\min}}=\min\left(  D_{\mathrm{a}},D_{\mathrm{b}}\right)  $.
As can be seen from Eq. (\ref{linear relative entropy S(rho)}), the linear
relative entropy $\sigma_{\mathrm{LR}}$ quantifies the correlation between
subsystems a and b \cite{Wolf_070502}. Note that $\sigma_{\mathrm{LR}}$\ is
generally bounded by $\sigma_{\mathrm{LR}}\in\left[  0,1\right]  $.

\textbf{Constraints} -- Let $\rho_{\mathrm{ME}}$ be the density operator of
the composed system, which maximizes the entropy $\sigma=-\operatorname{Tr}%
\left(  \rho\log\rho\right)  $, under the constraints that both $\rho
_{\mathrm{a}}$ and $\rho_{\mathrm{b}}$ are fixed (i.e. the Bloch vectors
$\mathbf{P}_{\mathrm{a}}$ of subsystem a, and $\mathbf{P}_{\mathrm{b}}$ of
subsystem b are fixed). The maximum entropy density operator $\rho
_{\mathrm{ME}}$ can be found using the Lagrange multipliers method, which
yields $\rho\dot{=}Z^{-1}\exp\left(  -\boldsymbol{\alpha}_{\mathrm{a}}%
\cdot\mathbf{\Lambda}_{\mathrm{a}}-\boldsymbol{\alpha}_{\mathrm{b}}%
\cdot\mathbf{\Lambda}_{\mathrm{b}}\right)  $, where both vectors
$\boldsymbol{\alpha
}_{\mathrm{a}}$ and $\boldsymbol{\alpha}_{\mathrm{b}}$\ are real, the
partition function $Z$ is given by $Z=\operatorname{Tr}\exp\left(
-\boldsymbol{\alpha
}_{\mathrm{a}}\cdot\mathbf{\Lambda}_{\mathrm{a}}%
-\boldsymbol{\alpha}_{\mathrm{b}}\cdot\mathbf{\Lambda}_{\mathrm{b}}\right)  $,
and the vectors $\mathbf{\Lambda}_{\mathrm{a}}$ and $\mathbf{\Lambda
}_{\mathrm{b}}$ of $D_{\mathrm{H}}\times D_{\mathrm{H}}$ matrices are given by
$\mathbf{\Lambda}_{\mathrm{a}}=\left(  G_{1,0},G_{2,0},\cdots,G_{D_{\mathrm{a}%
}^{2}-1,0}\right)  $ and $\mathbf{\Lambda}_{\mathrm{b}}=\left(  G_{0,1}%
,G_{0,2},\cdots,G_{0,D_{\mathrm{b}}^{2}-1}\right)  $
\cite{Jaynes_579,Presse_1115}. Note that $\rho_{\mathrm{ME}}$ is positive semi
definite, and that $\rho_{\mathrm{ME}}=\rho_{\mathrm{a}}\otimes\rho
_{\mathrm{b}}$ [see inequality (\ref{subadditivity})].

For the quantum equations of motions (\ref{modified master equation}) and
(\ref{SLE}), the constraints \cite{Maity_022202} that both $\rho_{\mathrm{a}}$
and $\rho_{\mathrm{b}}$ are fixed (for the case $\mathcal{H}=0$) are enforced
by applying the transformation%
\begin{equation}
\Theta^{{}}\rightarrow\Theta^{\prime}=\Theta+\sum_{a=0}^{d_{\mathrm{a}}^{2}%
}\sum_{b=0}^{d_{\mathrm{b}}^{2}}\eta_{ab}G_{a,b}\;.
\label{Theta -> Theta prime}%
\end{equation}
The Lagrange coefficients $\eta_{ab}$\ are determined from the requirement
that both $\rho_{\mathrm{a}}$ and $\rho_{\mathrm{b}}$ are fixed, which is
expressed as a set of $D_{\mathrm{H}}^{2}$ constraints given by
$\operatorname{Tr}\left(  \Omega\left(  \Theta^{\prime}\right)  G_{a,b}%
\right)  =\operatorname{Tr}\left(  \Omega\left(  \Theta\right)  G_{a,b}%
\right)  $, for $\left(  a=b=0\right)  \vee\left(  a\neq0\wedge b\neq0\right)
$, and $\operatorname{Tr}\left(  \Omega\left(  \Theta^{\prime}\right)
G_{a,b}\right)  =0$, for $\left(  a=0\wedge b\neq0\right)  \vee\left(
a\neq0\wedge b=0\right)  $ [the symbol $\wedge$\ denotes logical and, the
symbol $\vee$\ denotes logical or, and see Eq. (\ref{modified master equation}%
)]. Note that the Lagrange coefficients $\eta_{ab}$ can be extracted from the
constraints, provided that the mapping $\tilde{\Omega}\left(  X\right)  $ is invertible.

Consider the case where the operator $\Theta$ is given by
$\Theta=\gamma\left(  \log\rho+\log d\right)  $, where $\gamma$ is the rate of
disentanglement and $d$ is the Hilbert space dimensionality. Note that
$\Theta$ is positive semi--definite (see the second paragraph of appendix
\ref{AppK}). The spectral decomposition of the density operator $\rho$ is
expressed as $\rho=\sum_{n}r_{n}\left\vert \phi_{n}\right\rangle \left\langle
\phi_{n}\right\vert $, where $r_{n}\in\left[  0,1\right]  $ and $\sum_{n}%
r_{n}=1$. In the basis $\left\{  \left\vert \phi_{n}\right\rangle \right\}  $
of eigenvectors of $\rho$, the operator $\Omega\left(  \Theta\right)
=-\Theta\rho-\rho\Theta+2\left\langle \Theta\right\rangle \rho$ is given by%
\begin{equation}
\Omega\left(  \Theta\right)  =-2\gamma\sum_{n}r_{n}\left(  \log r_{n}%
-\left\langle \log\rho\right\rangle \right)  \left\vert \phi_{n}\right\rangle
\left\langle \phi_{n}\right\vert \;, \label{Omega(Theta) SD}%
\end{equation}
where $-\left\langle \log\rho\right\rangle =\sum_{n}f\left(  r_{n}\right)  $
is the entropy, and where $f\left(  r\right)  =-r\log r$.

Consider the case where $\rho$ is nearly pure $\rho\simeq\left\vert
\phi_{n_{0}}\right\rangle \left\langle \phi_{n_{0}}\right\vert $, i.e.
$r_{n_{0}}\simeq1$ for some integer $n_{0}$, and $r_{n}\simeq0$ for $n_{{}%
}\neq n_{0}$. Note that $\lim_{r\rightarrow0}f\left(  r\right)  =f\left(
1\right)  =0$, and thus $\Omega\left(  \Theta\right)  =0$ for any pure state.
The first derivative $f^{\prime}$ of the function $f$ is given by $f^{\prime
}=-\log r-1$. Note that, even though $\lim_{r\rightarrow0}f^{\prime}\left(
r\right)  =\infty$, the function $f\left(  r\right)  $ is continuous for
$r\in\left[  0,1\right]  $, which implies that the operator $\Omega\left(
\Theta\right)  $ is continuous. These observations suggest that for the case
$\mathcal{H}=0$, any pure state is a fixed point of the modified master
equation (\ref{modified master equation}), however, that fixed point is
unstable [provided that $\gamma>0$, see Eq. (\ref{Omega(Theta) SD})].

\begin{figure*}[ptb]
\begin{center}
\includegraphics[width=6.4in,keepaspectratio]{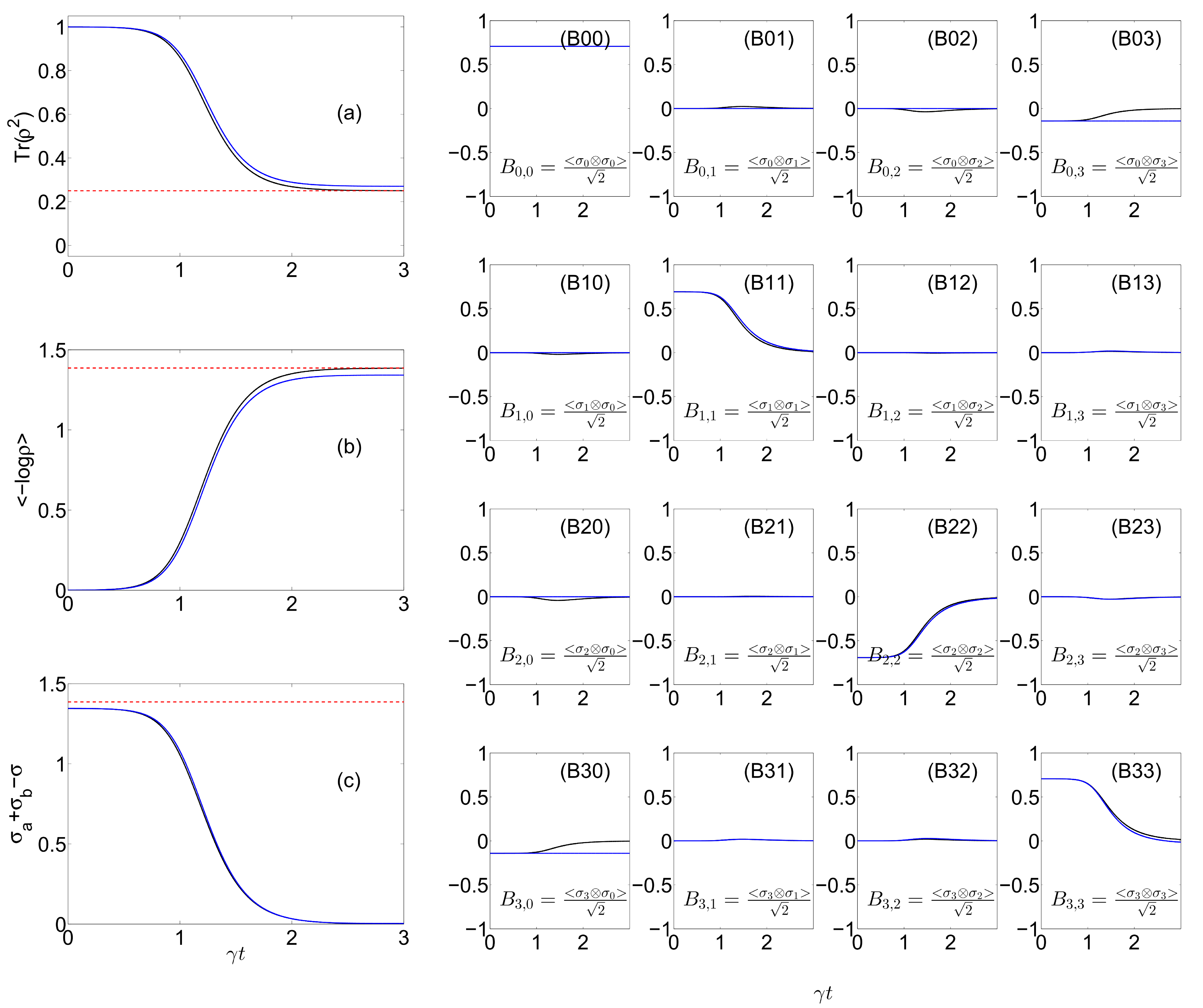}
\end{center}
\caption{{}Two spin 1/2 system. Plots generated without (with) applying the
constraints are black (blue) colored. The (a) purity $\operatorname{Tr}%
\rho^{2}$, (b) total entropy $\sigma=\left\langle -\log\rho\right\rangle $ and
(c) relative entropy $\sigma\left(  \rho\parallel\rho_{\mathrm{a}}\otimes
\rho_{\mathrm{b}}\right)  =\sigma_{\mathrm{a}}+\sigma_{\mathrm{b}}-\sigma_{{}%
}$ [see Eq. (8.223) of Ref. \cite{Buks_QMLN}] are plotted as a function of
normalized time $\gamma t$. The overlaid horizontal dotted red lines in (a),
(b) and (c) represents the values of $1/4$, $\log4$ and $\log4$, respectively.
The time dependency of the Bloch matrix element $B_{a,b}$ is shown in
($\mathrm{Bab}$), where $a\in\left\{  0,1,2,3\right\}  $ and $b\in\left\{
0,1,2,3\right\}  $. The $2\times2$ identity matrix is denoted by $\sigma_{0}$,
and $\sigma_{1}$, $\sigma_{2}$ and $\sigma_{3}$ are Pauli matrices.}%
\label{FigTwoQ}%
\end{figure*}

\begin{figure*}[ptb]
\begin{center}
\includegraphics[width=6.2in,keepaspectratio]{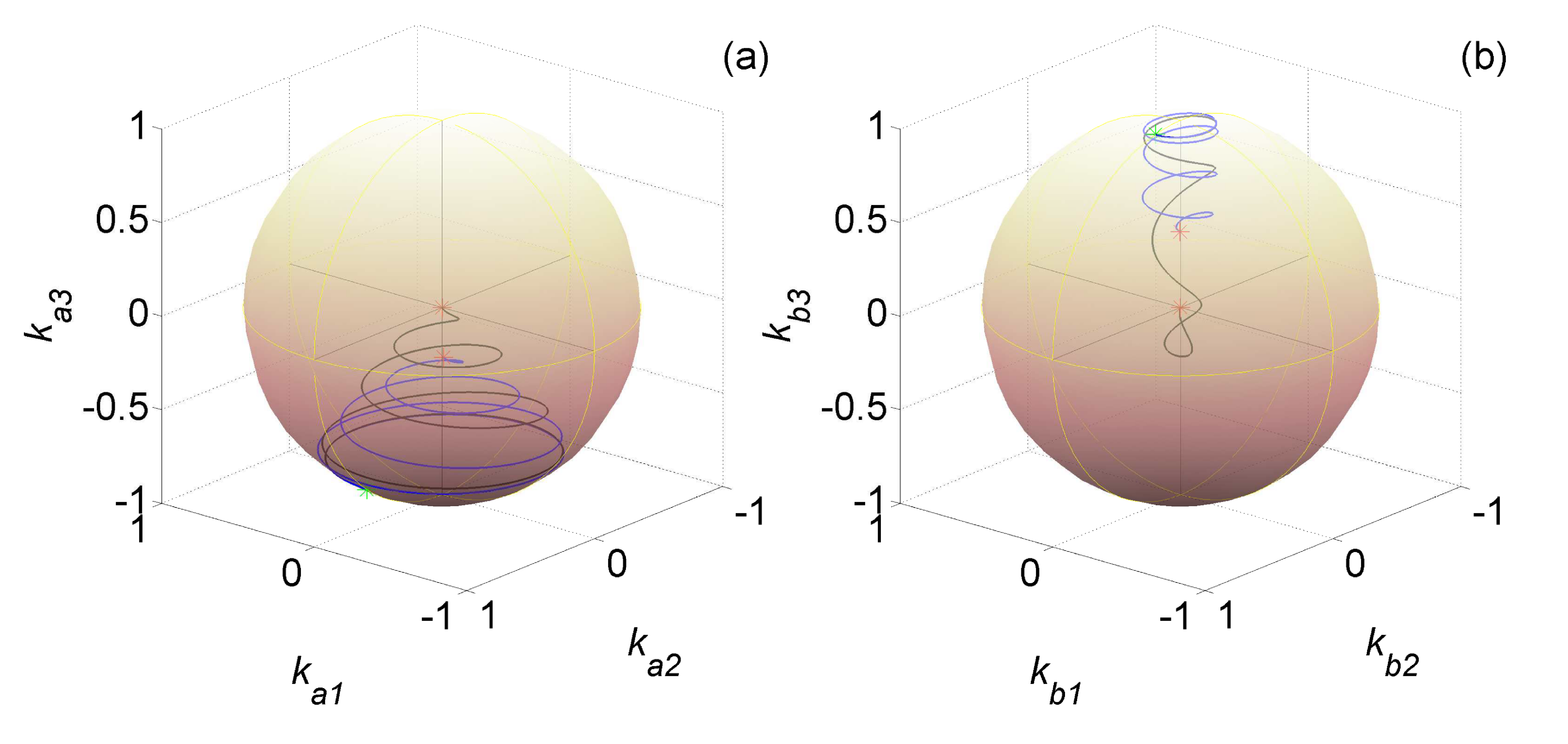}
\end{center}
\caption{{}Dipolar coupling. Time evolution of the single spin normalized
Bloch vectors $\mathbf{k}_{\mathrm{a}}$ and $\mathbf{k}_{\mathrm{b}}$ is shown
in (a) and (b), respectively. The color black (blue) is used to label the
plots that have been obtained without (with) applying the constraints. The
dipolar coupling coefficient is $\omega=100$, and the rate of disentanglement
is $\gamma=3$. Note that initially $\left\vert \mathbf{k}_{\mathrm{a}%
}\right\vert =\left\vert \mathbf{k}_{\mathrm{b}}\right\vert =1$ (see the
points labeled by a green $\times$ symbol), since the initial state is both
pure and fully disentangled.}%
\label{FigSzSz}%
\end{figure*}

\textbf{Two spin 1/2 system} -- The plots shown in Fig. \ref{FigTwoQ} and in
Fig. \ref{FigSzSz} demonstrate the disentanglement process for the case of a
system composed of two spin 1/2 particles. For Fig. \ref{FigTwoQ}, the
Hamiltonian vanishes (i.e. $\mathcal{H}=0$), and the initial state, which is
pure, is given by $\left\vert \psi\right\rangle \left\langle \psi\right\vert
$. In the Schmidt basis, the initial state vector $\left\vert \psi
\right\rangle $ is expressed as $\left\vert \psi\right\rangle =\sqrt
{p}\left\vert \uparrow\uparrow\right\rangle +\sqrt{1-p}\left\vert
\downarrow\downarrow\right\rangle $ [see Eq. (\ref{Schmidt decomposition})],
where $p\in\left[  0,1\right]  $ ($p=2/5$ for the plots shown in Fig.
\ref{FigTwoQ}). The state's time evolution if found by numerically integrating
the modified master equation (\ref{modified master equation}). The
nonlinear term is constructed using the operator $\Theta=\gamma\left(
\log\rho+\log4\right)  $.

The plots on the left side of Fig. \ref{FigTwoQ} display the time dependency
of (a) the purity $\operatorname{Tr}\rho^{2}$, (b) the total entropy
$\sigma=\left\langle -\log\rho\right\rangle $ and (c) the relative entropy
\cite{Wehrl_221,Vedral_2275,Bianconi_066001} $\sigma\left(  \rho\parallel
\rho_{\mathrm{a}}\otimes\rho_{\mathrm{b}}\right)  =\sigma_{\mathrm{a}}%
+\sigma_{\mathrm{b}}-\sigma_{{}}$ [see inequality (\ref{subadditivity}), and
note that generally $\sigma\left(  \rho^{\prime}\parallel\rho^{\prime\prime
}\right)  \equiv\operatorname{Tr}\left(  \rho^{\prime}\left(  \log\rho
^{\prime}-\log\rho^{\prime\prime}\right)  \right)  \geq0$].

For the plots that are black colored, no constraints are applied, whereas the
color blue is used to label the plots that are obtained by applying the
constraints [see Eq. (\ref{Theta -> Theta prime})]. The $4\times4$ block of
plots that are labeled by the capital letter B display the time dependency of
the $4\times4=16$ entries $B_{a,b}$ of the Bloch matrix $B$. Note that, for
the blue--colored plots (for which the constraints are applied), the single
spin Bloch vectors $\mathbf{P}_{\mathrm{a}}=\left(  B_{1,0},B_{2,0}%
,B_{3,0}\right)  $ and $\mathbf{P}_{\mathrm{b}}=\left(  B_{0,1},B_{0,2}%
,B_{0,3}\right)  $ are unaffected by the disentanglement process.
Nevertheless, as can be seen from Fig. \ref{FigTwoQ}(c), the efficiency of
disentanglement is nearly unaffected by the constraints.

The effect of dipolar coupling is demonstrated by the plots shown in Fig.
\ref{FigSzSz}, which depict the time evolution of the single spin normalized
Bloch vectors $\mathbf{k}_{\mathrm{a}} \equiv2^{1/2} \mathbf{P}_{\mathrm{a}}$
and $\mathbf{k}_{\mathrm{b}} \equiv2^{1/2} \mathbf{P}_{\mathrm{b}}$. The
initial state $\left\vert \psi\right\rangle $ at time $t=0$, which is assumed
to be both pure and fully disentangled, is labeled by a green $\times$ symbol,
whereas a red $\times$ symbol is used to label the final state. The dipolar
coupling is described by the Hamiltonian $\mathcal{H}$, which is given by
$\mathcal{H}=\omega\sigma_{3}\otimes\sigma_{3}$, where $\omega$ denotes the
dipolar coupling coefficient. Without applying the constraints, the system
evolves towards the fully mixed state, for which $\mathbf{k}_{\mathrm{a}%
}=\mathbf{k}_{\mathrm{b}}=0$ (see the red $\times$ symbols for the black
colored plots). On the other hand, finite normalized Bloch vectors
$\mathbf{k}_{\mathrm{a}}$ and $\mathbf{k}_{\mathrm{b}}$ are obtained in the
long time limit when the constraints are applied (see the red $\times$ symbols
for the blue colored plots). Note that in the long time limit both
$\mathbf{k}_{\mathrm{a}}$ and $\mathbf{k}_{\mathrm{b}}$ are parallel (or
anti--parallel) to the dipolar coupling direction (the $z$ axis). Thus the
time evolution generated by both the dipolar coupling $\sigma_{3}\otimes
\sigma_{3}$ and the constrained process of disentanglement mimics a
measurement of the spins' angular momentum $z$ component (i.e. $k_{\mathrm{a}%
3}$ and $k_{\mathrm{b}3}$).

\textbf{Discussion} -- The examples presented in Figs. \ref{FigTwoQ} and
\ref{FigSzSz} demonstrate that the conflict with the principle of causality
can be mitigated. However, the proposed formulation is based on
non--relativistic QM, and consequently, full reconciliation with causality is
seemingly unachievable within this framework (e.g. similarly to standard QM,
superluminal tunneling cannot be excluded). Nevertheless, the proposed
formulation allows incorporating unitary time evolution with the processes of
disentanglement and thermalization, and it can be used to derive an effective
model for some nonlinear effects in quantum systems.

Spontaneous disentanglement \cite{Buks_e00986} makes the collapse postulate of
QM redundant. Disentanglement has no effect on any product (i.e. disentangled)
state, thus, all predictions of standard QM are unchanged in the absence of
entanglement. For a multipartite system, disentanglement between any pair of
subsystems can be introduced. Disentanglement is invariant under any subsystem
unitary transformation, and it is applicable for both distinguishable and
indistinguishable particles \cite{Buks_630}. Recently, the hypothesis has been
experimentally tested using a spin resonator \cite{Buks_e00909}.

Further study is needed to explore alternative methods for
incorporating unitary time evolution with the process of thermalization. One
leading candidate for that is the Grabert master equation
\cite{Grabert_161,Ottinger_026121}. This master equation yields experimentally
testable predictions \cite{Ottinger_052119}. These predictions, similarly to
the ones that are derived from the spontaneous disentanglement hypothesis, are
experimentally falsifiable, since they are distinguishable from what is
derived from standard QM.

\textbf{Summary} -- In the current study, the conflict between the spontaneous
disentanglement hypothesis and the causality principle is explored. A
formulation of the hypothesis, which is based on the maximum entropy
principle, is proposed, and it is found that the conflict with the causality
principle can be mitigated by introducing constraints, which ensure that
subsystems' properties are unaffected by the process of disentanglement.

\textbf{Acknowledgments} -- Useful discussions with Diosi Lajos and Jasper van
Wezel are acknowledged.

\appendix

\section{Nonlinear Kraus operators}

\label{AppK}

Consider the case where, to first order in the time interval $\tau$, the
density operator $\rho$\ evolves according to%
\begin{equation}
\rho\left(  t+\tau\right)  =\sum_{k\in\left\{  0,1\right\}  }K_{k}^{{}}%
\rho\left(  t\right)  K_{k}^{\dag}+O\left(  \tau^{2}\right)  \;,
\label{rho(t+tau)}%
\end{equation}
where $K_{0}^{{}}=1-\left(  i\hbar^{-1}\mathcal{H}+\Theta\right)  \tau$ and
$K_{1}^{{}}=\sqrt{2\left\langle \Theta\right\rangle \tau}$\ are Kraus
operators
\cite{Elben_200501,Sergi_1350163,Brody_230405,Kaplan_055002,Geller_2200156},
which satisfy the averaged norm conservation condition $\left\langle
K_{0}^{\dag}K_{0}^{{}}+K_{1}^{\dag}K_{1}^{{}}\right\rangle =1+O\left(
\tau^{2}\right)  $ \cite{Daraban_048}, $\hbar$ is the reduced Planck's
constant, $\mathcal{H}^{{}}=\mathcal{H}^{\dag}$ is the system's Hamiltonian,
the positive semi--definite operator $\Theta$ is allowed to depend on $\rho$,
and $\left\langle \Theta\right\rangle =\operatorname{Tr}\left(  \Theta
\rho\right)  $. The expression for $\rho\left(  t+\tau\right)  $
(\ref{rho(t+tau)}), which implies that%
\begin{equation}
\frac{\rho\left(  t+\tau\right)  -\rho\left(  t\right)  }{\tau}=i\hbar
^{-1}\left[  \rho,\mathcal{H}\right]  +\Omega\left(  \Theta\right)  +O\left(
\tau\right)  \;, \label{rho(t+tau)-rho(t)}%
\end{equation}
where $\Omega\left(  \Theta\right)  =-\Theta\rho-\rho\Theta+2\left\langle
\Theta\right\rangle \rho$, yields the master equation
(\ref{modified master equation}) in the main text.

Note that the time evolution that is governed by the master equation
(\ref{modified master equation}) is invariant under the transformation
$\Theta^{\prime}\rightarrow\Theta^{\prime\prime}=\Theta^{\prime}+xI$, where
$x$ is real and $I$ is the identity operator [since $\Omega\left(
\Theta^{\prime\prime}\right)  =\Omega\left(  \Theta^{\prime}\right)  $
provided that $\operatorname{Tr}\rho=1$]. This transformation can be used to
map any given Hermitian operator $\Theta^{\prime}$ into a positive
semi--definite operator $\Theta$, provided that $\Theta^{\prime}$ is bounded.
For example, for the Hermitian operator $\Theta^{\prime}=\gamma\log\rho$,
where $\gamma$ is positive, the transformed operator $\Theta=\gamma\left(
\log\rho+I\log d\right)  $, where $d$ is the Hilbert space dimensionality has
an expectation value given by $\left\langle \Theta\right\rangle =\gamma\left(
\operatorname{Tr}\left(  \rho\log\rho\right)  +\log d\right)  $, provided that
$\operatorname{Tr}\rho=1$, and thus $0\leq\left\langle \Theta\right\rangle
\leq\gamma\log d$.

The time evolution of the purity $\mathrm{\operatorname{Tr}}\rho^{2}$ is
governed by [see Eq. (\ref{modified master equation})]%
\begin{align}
\frac{\mathrm{d\operatorname{Tr}}\rho^{2}}{\mathrm{d}t}  &
=\mathrm{\operatorname{Tr}}\left(  2\frac{\mathrm{d}\rho}{\mathrm{d}t}%
\rho\right) \nonumber\\
&  =4\mathrm{\operatorname{Tr}}\left(  \left(  \left\langle \Theta
\right\rangle -\Theta\right)  \rho^{2}\right)  \;.\nonumber\\
&  \label{(d/dt)Tr(rho^2)}%
\end{align}
Note that%
\begin{equation}
\frac{\operatorname{Tr}\Omega\left(  \Theta\right)  }{2}=\operatorname{Tr}%
\mathcal{D}_{\rho}\left(  \sqrt{\Theta}\right)  \;,
\end{equation}
provided that $\operatorname{Tr}\rho=1$, where the Lindbladian $\mathcal{D}%
_{\rho}\left(  X\right)  $ for an operator $X$\ is given by%
\begin{equation}
\mathcal{D}_{\rho}\left(  X^{{}}\right)  =X^{{}}\rho X^{\dag}-\frac{X^{\dag
}X^{{}}\rho+\rho X^{\dag}X^{{}}}{2}\;.
\end{equation}

The master equation (\ref{modified master equation}) can be obtained from a
stochastic Langevin--Schr\"{o}dinger equation for the state vector $\left\vert
\psi\right\rangle $ \cite{Grimaudo_033835,Kowalski_167955}. Consider an
equation for $\left\vert \psi\right\rangle $ having the form%
\begin{equation}
\frac{\mathrm{d}\left\vert \psi\right\rangle }{\mathrm{d}t}=\left(
-i\hbar^{-1}\mathcal{H}+\eta_{0}\xi\left(  t\right)  +\eta_{1}\right)
\left\vert \psi\right\rangle \;, \label{SLE eta}%
\end{equation}
where $\eta_{0}$ and $\eta_{1}$ are operators. The white noise term
$\xi\left(  t\right)  $ has a vanishing averaged value, i.e. $\overline
{\xi\left(  t\right)  }=0$, and a correlation function given by $\overline
{\xi^{{}}\left(  t^{\prime}\right)  \xi^{\ast}\left(  t^{\prime\prime}\right)
}=\delta\left(  t^{\prime}-t^{\prime\prime}\right)  $ \cite{Rahman_354}, where
overbar denotes time averaging. The Schr\"{o}dinger--Langevin equation
(\ref{SLE eta}) yields to first order in the time interval $\tau$%
\begin{align}
&  \left\vert \psi\left(  t+\tau\right)  \right\rangle \nonumber\\
&  =\left(  1+\tau\left(  -i\hbar^{-1}\mathcal{H}+\eta_{1}\right)  +\eta
_{0}I\right)  \left\vert \psi\right\rangle +O\left(  \tau^{2}\right)
\;,\nonumber\\
&  \label{|psi(t+tau>}%
\end{align}
where the Stratonovich integral $I$ is given by%
\begin{equation}
I=\int_{t}^{t+\tau}\mathrm{d}t^{\prime}\;\xi\left(  t^{\prime}\right)  \;.
\end{equation}
The following holds%
\begin{equation}
\overline{I}=\int_{t}^{t+\tau}\mathrm{d}t^{\prime}\;\overline{\xi\left(
t^{\prime}\right)  }=0\;,
\end{equation}
and%
\begin{equation}
\overline{I^{{}}I^{\ast}}=\int_{t}^{t+\tau}\mathrm{d}t^{\prime}\int
_{t}^{t+\tau}\mathrm{d}t^{\prime\prime}\;\overline{\xi^{{}}\left(  t^{\prime
}\right)  \xi^{\ast}\left(  t^{\prime\prime}\right)  }=\tau\;.
\end{equation}

The relation [see Eq. (\ref{|psi(t+tau>})]%
\begin{align}
&  \overline{\left\langle \psi\left(  t+\tau\right)  \right.  \left\vert
\psi\left(  t+\tau\right)  \right\rangle }\nonumber\\
&  =\left\langle \psi\left(  t\right)  \right.  \left\vert \psi\left(
t\right)  \right\rangle +\tau\left\langle \psi\left(  t\right)  \right\vert
\eta_{1}^{{}}+\eta_{1}^{\dag}+\eta_{0}^{\dag}\eta_{0}^{{}}\left\vert
\psi\left(  t\right)  \right\rangle \nonumber\\
&  +O\left(  \tau^{2}\right)  \;,\nonumber\\
&
\end{align}
implies that norm is conserved on average by Eq. (\ref{SLE eta}) provided that (it is
assumed that $\left\langle \psi\right.  \left\vert \psi\right\rangle =1$ at
time $t$)%
\begin{equation}
0=\left\langle \eta_{1}^{{}}+\eta_{1}^{\dag}+\eta_{0}^{\dag}\eta_{0}^{{}%
}\right\rangle \;,\label{norm conservation eta}%
\end{equation}
and the relation [see Eq. (\ref{|psi(t+tau>})]%
\begin{align}
&  \overline{\left\vert \psi\left(  t+\tau\right)  \right\rangle \left\langle
\psi\left(  t+\tau\right)  \right\vert }\nonumber\\
&  =\left\vert \psi\right\rangle \left\langle \psi\right\vert \nonumber\\
&  +\tau\left\vert \psi\right\rangle \left\langle \psi\right\vert \left(
i\hbar^{-1}\mathcal{H}+\eta_{1}^{\dag}\right)  +\tau\left(  -i\hbar
^{-1}\mathcal{H}+\eta_{1}^{{}}\right)  \left\vert \psi\right\rangle
\left\langle \psi\right\vert \nonumber\\
&  +\tau\eta_{0}^{{}}\left\vert \psi\right\rangle \left\langle \psi\right\vert
\eta_{0}^{\dag}+O\left(  \tau^{2}\right)  \;,\nonumber\\
&
\end{align}
implies that the Schr\"{o}dinger--Langevin equation (\ref{SLE eta}) yields Eq.
(\ref{rho(t+tau)-rho(t)}) provided that [recall that $\Omega\left(
\Theta\right)  =-\Theta\rho-\rho\Theta+2\left\langle \Theta\right\rangle \rho
$]%
\begin{equation}
\rho\eta_{1}^{\dag}+\eta_{1}^{{}}\rho+\eta_{0}^{{}}\rho\eta_{0}^{\dag}%
=\Omega\left(  \Theta\right)  \;.\label{eta Omega}%
\end{equation}

As an example, both conditions (\ref{norm conservation eta}) and
(\ref{eta Omega}) are satisfied by the operators $\eta_{0}=\sqrt{2\left(
1-\zeta\right)  \left\langle \Theta\right\rangle }$ and $-\eta_{1}%
=\Theta-\zeta\left\langle \Theta\right\rangle $, where $\zeta\in\left[
0,1\right]  $. For the case $\zeta=0$. i.e. for $\eta_{0}=\sqrt{2\left\langle
\Theta\right\rangle }$ and $-\eta_{1}=\Theta$, the Schr\"{o}dinger--Langevin
equation (\ref{SLE eta}) becomes%
\begin{equation}
\frac{\mathrm{d}\left\vert \psi\right\rangle }{\mathrm{d}t}=\left(
-i\hbar^{-1}\mathcal{H}+\sqrt{2\left\langle \Theta\right\rangle }\xi\left(
t\right)  -\Theta\right)  \left\vert \psi\right\rangle \;.\label{SLE}%
\end{equation}
Note that for the case $\zeta=1$. i.e. for $\eta_{0}=0$ and $-\eta_{1}%
=\Theta-\left\langle \Theta\right\rangle $, the equation (\ref{SLE eta}) for
$\left\vert \psi\right\rangle $ becomes deterministic (no noise term)%
\begin{equation}
\frac{\mathrm{d}\left\vert \psi\right\rangle }{\mathrm{d}t}=\left(
-i\hbar^{-1}\mathcal{H}-\Theta+\left\langle \Theta\right\rangle \right)
\left\vert \psi\right\rangle \;.
\end{equation}

\bibliographystyle{ieeepes}
\bibliography{acompat,Eyal_Bib}

\end{document}